\documentclass[10pt,preprint]{aastex}

\shorttitle{Polarization of NGC 1068}
\shortauthors{Simpson et al.}
 
\begin{document}

\title{{\it HST} NICMOS Observations of the Polarization of NGC 1068\footnote
{Based on observations made with the NASA/ESA {\it Hubble Space Telescope}, 
obtained at the Space Telescope Science Institute, which is operated by 
the Association of Universities for Research in Astronomy, Inc., 
under NASA contract NAS 5-26555. These observations are associated with GTO proposal \#7212,
PI: E. F. Erickson.}
}

\author{Janet P. Simpson\altaffilmark{2}, Sean W. J. Colgan, Edwin F. Erickson}
\affil{Astrophysics Branch, MS 245-6, NASA Ames Research Center, Moffett Field, CA 94035-1000}
\email{simpson@cygnus.arc.nasa.gov; colgan@cygnus.arc.nasa.gov; erickson@cygnus.arc.nasa.gov}

\author{Dean C. Hines}
\affil{Steward Observatory, University of Arizona, Tucson, AZ 85721}
\email{dhines@as.arizona.edu}

\author{A. S. B. Schultz}
\affil{School of Physics, University of New South Wales, Sydney, NSW 2052, Australia}
\email{schultz@phys.unsw.edu.au}

\and

\author{Susan R. Trammell}
\affil{Department of Physics, University of North Carolina, Charlotte, NC 28223}
\email{srtramme@uncc.edu}

\altaffiltext{2}{SETI Institute}

\begin{abstract}
We have observed the polarized light at 2 \micron\ in the center of NGC 1068 with {\it HST}
NICMOS Camera 2.
The nucleus is dominated by a bright, unresolved source, 
polarized at a level of $6.0 \pm 1.2$\% 
with a position angle of $122^\circ \pm 1.5^\circ$.
There are two polarized lobes extending up to $8''$ northeast and southwest of the nucleus.
The polarized flux in both lobes is quite clumpy, with the maximum polarization occurring
in the southwest lobe at a level of 17\% when smoothed to $0.23''$ resolution.
The perpendiculars to the polarization vectors in these two lobes point back to the
intense unresolved nuclear source to within one $0.076''$ Camera 2 pixel, 
thereby confirming that this is the illuminating source of the scattered light
and therefore the probable AGN central engine.
Whereas the polarization of the nucleus is probably caused by dichroic absorption,
the polarization in the lobes is almost certainly caused by scattering, 
with very little contribution from dichroic absorption.
Features in the polarized lobes include a gap at a distance of about $1''$ from the nucleus
toward the southwest lobe and a ``knot'' of emission about $5''$ northeast of the nucleus.
Both features had been discussed by ground-based observers, but they are much better defined
with the high spatial resolution of NICMOS.
The northeast knot may be the side of a molecular cloud that is facing the nucleus,
which cloud may be preventing the expansion of the northeast radio lobe at the head
of the radio synchrotron-radiation-emitting jet.

We also report the presence of two ghosts in the Camera 2 polarizers.
These had not been detected previously (Hines et al. 2000) because
they are relatively faint and require observations of a source with a large
dynamic range.

\end{abstract}

\keywords{galaxies: individual (NGC 1068) --- galaxies: Seyfert  --- polarization}

\section{Introduction}

NGC 1068 is the closest (14.4 Mpc for H$_0 = 75$ km s$^{-1}$ Mpc$^{-1}$, 72 pc arcsec$^{-1}$) 
Seyfert 2 galaxy; 
thus it has been studied extensively at all wavelengths
(see the summary of the literature by Bland-Hawthorn et al. 1997a).
The barred spiral galaxy is $\sim 7'$ in diameter with the northern rotation axis
apparently tipped towards the earth at an angle of about $40^\circ$
(Schinnerer et al. 2000; Bland-Hawthorn et al. 1997a and references therein).
A cone of polarized (scattered) optical and ultraviolet (UV) light
(e.g., Scarrott et al. 1991; Miller, Goodrich, \& Mathews 1991; 
Capetti et al. 1995a, 1995b)
and co-spatial [\ion{O}{3}] (Pogge 1988; Bruhweiler et al. 2001) 
and [\ion{N}{2}] (Cecil, Bland, \& Tully 1990) emission 
fan out from the nucleus to the northeast at a position angle of $\sim 31^\circ$).
Radio jets emanate from a bright radio ``core'' along a similar position angle 
(Wilson \& Ulvestad 1983), but only a faint [\ion{O}{3}] counterpart
is observed toward the southwest (Unger et al. 1992)
and very little UV/optical polarized light (Scarrott et al. 1991). 
The faintness of the southwest cone suggests that it lies behind the plane of the
galaxy and is extinguished by the dust in the plane 
(e.g., Gallimore et al. 1994; Bland-Hawthorn et al. 1997b),
whereas it is thought that the brighter northeast cone emerges from the nucleus
in front of the galaxy plane (e.g., Kishimoto 1999; Bruhweiler et al. 2001).
Broad permitted lines including \ion{Fe}{2} are observed in the spectrum
measured in polarized light, 
confirming that this proto-type Seyfert 2 (narrow-lined) galaxy
harbors a Seyfert 1 (broad-lined) nucleus 
(Antonucci \& Miller 1985; Antonucci, Hurt, \& Miller 1994; Miller, Goodrich, \& Mathews 1991;
Inglis et al. 1995; Alexander et al. 1999)
that is hidden from our direct view by a geometrically and optically thick dusty region.
Light from the nucleus ionizes the clouds of the narrow line region;
moreover, ionizing and UV photons from the nucleus are scattered by electrons 
into the more extended, diffuse ionized gas of the galaxy disk and halo 
(Bland-Hawthorn, Sokolowski, \& Cecil 1991;
Sokolowski, Bland-Hawthorn, \& Cecil 1991; Neff et al. 1994).
The light is also scattered (thus polarized) by electrons and dust into our line of sight.

The nucleus of the galaxy is not well defined at UV and visible wavelengths;
instead, high resolution UV and visible imaging with the {\it Hubble Space Telescope} 
({\it HST}) only shows a number of bright clouds and filaments 
(Evans et al. 1991; Macchetto et al. 1994; Bruhweiler et al. 2001).
There have been a number of attempts to
estimate its location from the direction of
the UV polarization vectors measured with {\it HST}
(e.g., Capetti et al. 1995a, 1995b; Kishimoto 1999).
On the other hand, near-infrared (NIR) and mid-infrared images 
(Braatz et al. 1993; Marco, Alloin, \& Beuzit 1997; Thatte et al. 1997; 
Rouan et al. 1998; Weinberger, Neugebauer, \& Matthews 1999; Bock et al. 2000)
show an intense unresolved source.
Thompson et al. (2001), by correlating the optical H$\alpha$ and NIR P$\alpha$ images
from {\it HST},
determined that the NIR point source lies at the location of the nucleus 
(to within the errors of measurement) as determined by Kishimoto (1999).
These locations can be seen in Figure 7 and Table 2 of Kishimoto (1999),
where he plots the nucleus position determined from the UV polarization measurements
on the {\it HST} F501N [\ion{O}{3}] image.
The position (with uncertainty of $\sim 0.04''$) 
of Thompson et al. is consistent with both Kishimoto's nucleus position
and with Cloud B of Evans et al. (1991), also plotted on the figure.
The radio source S1 is thought to be the location
of the AGN nucleus and accretion disk 
because of the high velocities of its H$_2$O masers emanating 
from an apparently rotating, 4 pc diameter disk 
(Gallimore et al. 1996b, 2001; Greenhill \& Gwinn 1997)
with interior hot thermal emission (Muxlow et al. 1996; Roy et al. 1998).
The coordinates of source S1 agree with the positions of Cloud B and the nuclear
position plotted by Kishimoto (1999) within the absolute uncertainty of $0.08''$
of the optical position of the galaxy (Capetti, Macchetto, \& Lattanzi 1997b).

If the intense NIR unresolved source is really the galaxy and AGN nucleus,
its light should be scattered and polarized with a centrosymmetric polarization
pattern. 
Moreover, since NIR wavelengths are less affected by extinction than 
visible or UV wavelengths, the morphology of the scattering region
can be better determined through measurement of the NIR polarized light.
Ground-based measurements of the NIR polarization in $\gtrsim 1''$ to $2''$ resolution
have been made by Young et al. (1996), Packham et al. (1997), and Lumsden et al. (1999).
They show that the polarization
is not consistent with scattering in the immediate vicinity of the nucleus;
Young et al., Packham et al., and Lumsden et al. all suggest that there is
additional polarization caused by the light from the nucleus being absorbed 
by aligned dust grains in the AGN torus.
Young et al. and Lumsden et al. also detect a dip in the polarized emission 
$\sim 1''$ south of the nucleus
and all three groups see a southwest lobe of polarized emission 
that is not observed at visible wavelengths.

In this paper we report on 2 \micron\ polarization observations taken by the
Near Infrared Camera and Multi-Object Spectrometer (NICMOS) on {\it HST}.
With the $0.2''$ spatial resolution of NICMOS, we see the unresolved polarized source
that must be the nucleus and numerous scattering clouds in the NE and SW lobes.
Section 2 describes the observations, section 3 the results and discussion, 
and section 4 summarizes the results.

\section{Observations and Data Reduction}

NGC 1068 was observed in two visits on 11 October and 15 November, 1997,
with the telescope rolled such that 
NICMOS Camera 2 was oriented at $63.519^\circ$ and $154.519^\circ$
East of North, respectively.
Integrations were taken at 4 dither positions separated by 10.5 pixels ($0.78''$) 
in each of the three Camera 2 polarizing filters, POL0L, POL120L, and POL240L,
which cover a 1.9 to 2.1 \micron\ bandpass.
The detector array was read out in MULTIACCUM mode with sample sequence 
STEP16 to accumulate total times of 64 or 80 seconds per integration for visit 1
and 128 seconds for visit 2.
We were able to recover the full dynamic range in this observation
even though the nucleus saturates in 8 or 16 s because the design 
of the STEP16 MULTIACCUM sequence incorporates very short dwell times
for the first few read-outs.

The raw data were calibrated using CALNICA (v. 3.3) as implemented by STScI in STSDAS v2.2.
We included the CALNICA task BIASEQ but not the tasks PEDSKY or PEDSUB
because there is no blank sky in the $19.4''$ images.
The reference files used were the most recent Institute flats and non-linearity
files (of January, 2001), but the dark files were provided by 
M. Rieke (2000, private communication).
Each reduced image required additional corrections for quadrant bias
and half the images needed slight adjustments for residual shading.
Since images requiring no shading adjustment occurred during both visits, using 
the STScI temperature sensitive darks would not have negated the need for adjustments.
These corrections were very small, $\lesssim 0.05$ counts s$^{-1}$ whereas the nucleus center
is $\sim~3600$ counts s$^{-1}$ and the image edges $\sim~0.5$ counts s$^{-1}$.
Although the need for such corrections is not noticeable in the total intensity images,
it is obvious in quantities that incorporate a difference image, such as the polarized flux.
 
For each polarizing filter, we aligned and shifted the four dither positions 
by centroiding the unresolved nucleus and then median combined 
them to remove bad pixels
using the IDL program, IDP3\footnote{http://nicmos.as.arizona.edu}.
We used the equations of Hines, Schmidt, \& Schneider (2000) to calculate
the total flux ($F$), percentage polarization ($p$), polarized flux ($p \times F$), 
and position angle for each visit;
we calculated statistical errors using the equations of Sparks \& Axon (1999)
and the estimated errors on the flux computed by CALNICA.
The original pixel scale was $0.0760''$ by $0.0753''$ per pixel;
the pixels were rectified and smoothed with a $3\times3$ smoothing function 
(i.e., to approximately the {\it HST} spatial resolution of $0.2''$ at 2.0 \micron) 
in the data reduction process to achieve higher signal/noise.

The total intensity, fractional polarization, and polarized flux images 
from the two visits are plotted in Figure 1. 
Two ghosts at pixel positions (23,35) and (8,20) 
were discovered in the POL0L and POL120L filters, respectively, and are marked in the images.
Although the total fluxes for the two visits are very similar,
the percentage polarization and polarized flux differ by 2--3 percent
with visit 2 generally showing less polarization. 
The cause of the difference in polarization is not known.
Because the differences between the visits in the percentage polarization and polarized flux 
are generally larger than the calculated statistical errors, 
we estimated the actual uncertainties (which we quote in later sections) 
as equal to 0.5 times the differences between the visits.

\section{Results and Discussion}

\subsection{Nucleus}

In Figure 1 the nucleus of NGC 1068 appears as an intense, unresolved source
convolved with the {\it HST} point spread function (PSF)
in both the total flux and the polarized flux images.
Since we do not have a real-time example of the PSF from a star, we calculated
the PSF for each of the visits using the program 
TinyTIM\footnote{http://www.stsci.edu/software/tinytim} (Krist \& Hook 1997, 1999).
The nuclear region with contours of the TinyTIM PSF is plotted in Figure 2.
It is not possible to remove the effects of the PSF by subtracting the calculated PSF 
from the observed nucleus ---
numerous blobs at the locations of PSF diffraction rings and spikes remain
no matter what PSF scaling was used.
In fact, one can see in Figure 2 that although there is general agreement between
the calculated PSF contours and the polarized flux, there are also many regions
that do not have good enough correspondence between the calculated PSF
and the observed blobs for complete PSF subtraction.
As a result, we can only say that there is a hint of additional structure 
within the first Airy ring surrounding the nucleus.
It is possible that the nucleus is slightly extended in the north-south direction
as was suggested by Weinberger et al. (1999), Rouan et al. (1998), and Thatte et al. (1997).
It is interesting that apparently
the radio jet is initially emitted in approximately a north-south direction
and is diverted to the $\sim 30^o$ position angle
$0.3''$ north of the nuclear source S1 (Gallimore, Baum, \& O'Dea 1996a;
Muxlow et al. 1996).

The nuclear polarization, averaged over the central $3 \times 3$ pixels, is $6.0 \pm 1.2$\% 
and the position angle is $122^\circ \pm 1.5^\circ$, 
perpendicular to the eventual direction of the radio jet but not its initial direction.
This percentage polarization is larger than that measured by Packham et al. (1997) 
or Lumsden et al (1999), probably because their measurements were diluted by
the non-polarized flux from the galaxy core seen in their larger beams.
The measured position angles are very similar.

Lumsden et al. (1997) studied their polarization position angles to check for
the centrosymmetric pattern that would indicate scattering of light originating
at a well defined location (e.g., the nucleus).
They found that the NIR polarization vectors in the vicinity of the nucleus 
are not consistent with scattering of nuclear light only;
instead they suggested that the nuclear emission is polarized by 
dichroic absorption as a result of passing through aligned dust grains.
However, their beams were typically $2''$ or larger and thus could include
non-nuclear light as well as nuclear.
Alternatively, the nuclear polarized light could include light backscattered 
from the cones above the nucleus and scattered a second time on the
accretion disk (Fischer, Henning, \& Yorke 1996; Schinnerer et al. 2000).

In order to estimate what fraction of the nuclear polarized light is
coming from the point source, we measured the nuclear fluxes 
and the calculated PSF in various apertures
and corrected the fluxes by dividing by the PSF fraction in the same apertures.
These results are given in Table 1.
We see that the polarized flux 
from the nucleus of NGC 1068 completely dominates the polarized flux 
that would be seen in any $1'' - 2''$ aperture such as were used in the ground-based 
polarization measurements; 
contributions from off-nuclear scattered light would be barely detectable.
Only a small fraction of the polarized flux measured within a $2''$ aperture 
is not flux from the essentially unresolved nucleus (subtracting the PSF-corrected
polarized flux measured in the $0.209''$ radius aperture from that measured in the $1.00''$
radius aperture gives $3.4 \pm 0.2$ mJy).

The polarization vectors are plotted on a contour map of the polarized flux in Figure 3.
Since the vectors describing the vibration of the electric field in the sky plane
are perpendicular to the projected line between the illuminating source
and the (optically thin) scatterers, one can locate the illuminator 
by computing the intersection of the lines perpendicular 
to the polarization position angle vectors
(e.g., Lumsden et al. 1999; Kishimoto 1999; Capetti et al. 1995a).
The procedure is to pick a pixel, assume it is the center,
calculate the vector angle to every other pixel, and subtract 
the perpendicular polarization angle at those pixels.
The means of the differences as a function of position angle from the assumed center 
are a minimum when the assumed center pixel 
is located at the illuminating source of the scattered light.
For the NICMOS polarization vectors for regions greater than $1''$ from the peak, 
the polarization vectors point to the intense peak within $< 1$ pixel.
This confirms that the illuminating source of the scattered light is the same at 2 \micron\ 
as the illuminating source estimated from optical polarization measurements 
(e.g., Kishimoto 1999), given the alignment of the optical and Paschen-$\alpha$
images of Thompson et al. (2001).

\subsection{Inner Two Arcseconds}

Even though most of the brighter emission blobs in the central $2''$ are 
due to the complex diffraction pattern from the bright central point source (i.e., the PSF), 
it is clear that there is some polarized
emission extending along a ridge $\lesssim 1''$ north and south of the nucleus 
(Figs. 2 and 4).
This appears to be the structure imaged
by Thompson et al. (2001) in P$\alpha$ (their Fig. 1c)
and may include some of cloud C of Evans et al. (1991).
The other clouds seen in the inner $1''$ in UV polarization 
(e.g., Capetti et al. 1995a, 1995b) are obscured by the PSF
and there is no good correspondence with the other UV polarization structures 
in the central $2''$.
The emission spot described by Thompson et al. (2001) is not 
bright in polarized light.

Young et al. (1996) and Lumsden et al. (1999) noticed that there is a dip
in the polarized flux about $1''$ south of the nucleus in their J and H images
but not in their K band image.
This dip is clearly present in our 2 \micron\ NICMOS image
(see especially the fractional polarization images in Fig. 1), although it is not
clear whether it is a completely empty gap or whether there are some clouds
of scatterers present (the fact that the observed polarized flux blobs do not completely
correspond to the TinyTIM-calculated PSF is to be expected 
and does not necessarily indicate additional scattering sources in the region).
Young et al. (1996) suggested that the dip is caused by obscuration from
the dusty torus surrounding the AGN; 
this torus would have to have a size $\sim 200$ pc, which is substantially larger
than the few-parsec-sized canonical AGN torus (see Antonucci 1993 and references therein)
that directs the radiation into the observed scattering cones.
Another possible source of the obscuring dust is the dense cloud or nuclear ring 
(diameter $\sim 5''$) seen in CO by Schinnerer et al. (2000).
Schinnerer et al. estimate this ring to have a mass of $5 \times 10^7$ M$\sun$
 centered on the nucleus.
Using the interstellar ratio of $A_V$/hydrogen column density of
Bohlin, Savage, \& Drake (1978), we estimate values of $A_V$ for this cloud
ranging from 6 to 15.
Although this cloud could provide the small dip seen by Young et al. and Lumsden et al.,
it would not provide enough extinction to produce a gap as empty as
the polarization gap in Fig. 1 appears to be.
We suggest that there is a real dearth of scatterers in the band $1''$ south
of the nucleus.

\subsection{Southwest and Northeast Lobes}

The southwest and northeast lobes (see Fig. 4 for nomenclature) 
are seen to be highly clumped, with locally
very strong polarization of 
17\% when smoothed with a $3 \times 3$ smoothing to $0.23''$ per smoothed pixel 
(the NICMOS resolution is $0.2''$) and 20\% in a single pixel.
The southwest lobe is quite faint at UV and visible wavelengths;
the reason probably is that it lies behind the plane of the host galaxy and
is extinguished by interstellar dust.
Observations of the H$\alpha$/H$\beta$ ratio by Bruhweiler et al. (2001) 
show that the northeast lobe seen in the ionized lines suffers very little extinction but 
the southwest lobe has H$\alpha$/H$\beta \gtrsim 6$ 
(the Case B recombination line ratio is 2.87).
This implies $A_V \gtrsim 2$ mag for ionized hydrogen clouds to the southwest 
of the nucleus.

In addition to inferring the location of the main illuminating source,
one can also use the polarization vectors to infer whether there is 
any additional illuminating source besides the strong intensity peak 
(which is very close to the illuminating source, if not the actual source), 
or whether there is additional polarization from dichroic absorption
by aligned interstellar dust grains in the galaxy.
To do this, we measured the deviations of the perpendiculars to the polarization vectors
from the vectors to the bright nucleus for regions not containing PSF contributions 
from the nucleus.
Most of the largest deviations from zero occur where the intensity is less than
the local maximum, in other words, in the regions of larger uncertainty.
The only locations with large deviations that are consistent between the visits 
are at the extreme west edges of the northern and southern lobes. 
The uncertainty, though, is of similar size to the deviation, about 2 degrees
whereas the deviation is about 4 degrees.
We conclude that there is no strong evidence of any additional polarization component
in the regions $1'' - 4''$ from the nucleus from dichroic absorption of grains
aligned by a magnetic field oriented other than perpendicular to the line
joining the nucleus and the grain location in the galaxy. 
However, since magnetic fields in galaxies, including NGC 1068 (Scarrott et al. 1991),
tend to follow the spiral arms, it is likely that any dichroic absorption
would produce polarization with the same orientation as that produced 
by scattering of light emitted by the nucleus.
On the other hand, we find that there is no correlation 
of the bands of polarized flux in the southwest lobe
and regions of high H$\alpha$/H$\beta$ (large extinction) 
in Figure 3 of Bruhweiler et al. (2001). 
Thus, dichroic absorption is probably not the main contributor to the high polarization 
that we see in the southwest lobe in spite of the high optical extinction
with the possible exception of the extreme southwest clump, where Bruhweiler et al. (2001)
saw no appreciable H$\alpha$ emission.
Most likely the higher polarized intensity in the southwest cone compared to the northeast cone
is that it contains a higher density of dust (larger filling factor). 

Jets emitting synchrotron radiation have been observed emanating from the nucleus by
Wilson \& Ulvestad (1983, 1987).
In Fig. 4, we plot the 4.9 GHz radio contours on top of our combined polarized flux image.
The jets appear to pass through the regions containing the scattering material
without spreading out; only after they are beyond the regions exhibiting
polarized emission do they spread out into extended lobes.
This is the first observed correlation of some aspect of the south radio lobe
with any optical, IR, or X-ray feature.
One possibility is that the scattering material in the NE and SW lobes seen in NIR polarization
is affecting the radio jets and the formation of the radio lobes;
another is that the jets have pushed material into the dense dust clouds 
that are visible in scattered light.

\subsection{Northeast Knot}

The northeast ``knot'' was first identified by Miller et al. (1991) as a region 
or knot of high visible polarization about $4.7''$ northeast of the nucleus.
An enhancement in the polarized emission is also apparent in the NIR images
of Young et al. (1996), Packham et al. (1997), and Lumsden et al. (1999).
However, this region does not stand out in the optical ionized line images of
Bruhweiler et al. (2001): ionized gas is present but extended over a 
much larger region than the northeast knot, the excitation from the [\ion{O}{3}]/H$\beta$ ratio
is much lower right at the position of the knot than the surrounding gas, and
the extinction inferred from the H$\alpha$/H$\beta$ ratio is only moderate.

The northeast knot is intriguing because it is clearly on the outside edge of
the NE radio lobe observed by Wilson \& Ulvestad (1983, 1987). In fact, it looks like
the jet is forced to go past the cloud/knot and cannot penetrate it, as though the
cloud is a massive obstruction in the path of the radio lobe.
There is a fainter cloud of polarized emission immediately north of the radio lobe
as well.
Wilson \& Ulvestad (1987) suggested that the radio polarization morphology and
limb brightening of the NE radio lobe are caused by the high energy particles
of the lobe encountering the interstellar medium of the galaxy in 
a radiative bow shock.
Schinnerer et al. (2000) measured $^{12}$CO (1--0) emission about $5''$ north
of the nucleus.
Their largest cloud is just east of the location of the northeast knot ---
we suggest that the northeast knot and the north cloud are the edges 
of this cloud that is facing the AGN nucleus.
This, then, would be the cloud that is confining the NE radio lobe.
Miller et al. (1991) also suggested that the northeast knot has
a direct view of the nucleus and its ionizing radiation.

\subsection{Electron Scattering vs. Dust Scattering}

It has been thought for some time that the ultraviolet and visible polarized
flux in the nucleus of NGC 1068 is due to scattering of light by
electrons in the vicinity of the AGN central engine 
(Miller et al. 1991; Antonucci 1993; Young et al. 1995).
Even though scattering by dust is more efficient than electron scattering, 
electron scattering is suggested for the nuclear region because
of the constancy of percentage polarization between UV and visible wavelengths, 
because the X-rays appear to be scattered in the same way as the visible
polarized flux (Antonucci 1993), and
because the H$\beta$ line shows evidence of broadening from scattering
by high temperature electrons (Miller et al. 1991).
The {\it Chandra} X-ray images (Young, Wilson, \& Shopbell 2001)
show good correlation with [\ion{O}{3}] and H$\alpha$ images 
(Capetti, Axon, \& Macchetto 1997a; Macchetto et al. 1994; Bruhweiler et al. 2001), 
which are similar to the {\it HST} UV polarized intensity images (Capetti et al. 1995a, 1995b).

If the polarized flux from the unresolved nucleus is due to dichroic absorption,
as seems likely,
then we should compare the non-nuclear NIR polarized flux to the 
visible polarized flux to investigate whether both are due to electron scattering.
Inglis et al. (1995) measured the polarized spectrum from 0.43 to 0.69 \micron\ 
in a $1.3'' \times 2.7''$ slit
at a number of positions, including positions $2.5''$ and $5.0''$ NE of the nucleus.
In the position $2.5''$ NE of the nucleus (approximately the NE lobe),
their polarized flux decreased from about 0.4 mJy to about 0.2 mJy over this
spectral range.
Our measured polarized flux in the same aperture is $\gtrsim 0.54$ mJy
after we subtract 0.15 mJy for the contribution from the diffraction spikes from
the nucleus, as computed using TinyTIM
(the actual peak of the NE lobe is a little further to the West and has larger flux).
The position $5.0''$ NE of the nucleus is that of the northeast knot.
Here Inglis et al. (1995) found that the polarized flux decreased 
from about 0.35 mJy at 0.45 \micron\ to about 0.25 mJy at 0.69 \micron.
In a similar aperture we measured the polarized flux to be
0.65 mJy in the northeast knot at 2.0 \micron.
For both positions the ratio of the polarized flux at 2.0 \micron\ divided 
by the polarized flux at 0.6 \micron\ is the same, $\sim 2.2$.
We conclude that the scattering is produced by the same process at both positions.

The possibilities for the large increase in scattered flux from visible to NIR wavelengths 
at these two non-nuclear positions are:
1) The intrinsic NGC 1068 Seyfert 1 AGN spectrum increases rapidly between
visible and NIR wavelengths, even more so than is typical 
(Ward et al. 1987 observed that  the fluxes, $f_\nu$, of bare, unreddened Seyfert 1 galaxy nuclei increase from visible to NIR wavelengths even though they are flat or increase
from visible to UV wavelengths). 
Such an increase in the nuclear flux could be due to an increasing contribution from the
very hot dust immediately surrounding the central engine.
2) The polarized flux seen at visible wavelengths suffers substantial extinction
such that the apparent flux ratios of NIR/visible are much larger than 
the intrinsic ratios. If this were the case, though, there should be almost
no UV polarized flux and the slowly decreasing polarized flux 
as a function of visible wavelength (Miller et al. 1991; Inglis et al. 1995) 
would not be observed. 
3) There are systematic effects such that NICMOS measures too much polarized flux. 
Overall, however, our nuclear photometry at 2.0 \micron\ 
is consistent with other NIR measurements 
(Weinberger et al. 1999; Lumsden et al. 1999; Packham et al. 1997), given the
steep rise in nuclear flux between the H and K bands,
and thus we do not consider this possibility to be likely.

For the northeast knot, 
Miller et al. (1991) suggested that the polarization was caused by dust scattering because 
the percentage polarization decreases rapidly to long wavelengths over the
observed wavelength range of 0.33 to 0.65 \micron\ 
and because the polarized H$\beta$ line is narrower than the 
nuclear polarized H$\beta$ line.
An additional reason for rejecting electron scattering as the cause
of the polarization in the northeast knot is that
the northeast knot is not particularly prominent in the Chandra image (Young et al. 2001),
whereas electron scattering is suggested as dominating 
the nuclear polarized flux because of the similarity of appearance of the nucleus in 
UV and X-ray images.
If the scattering is also caused by dust at 2 \micron, 
the nuclear source flux would have to increase
by much more than a factor of 2.2 since the scattering cross section for interstellar dust
decreases by a factor of order 20 from 0.6 \micron\ to 2.0 \micron.
Inglis et al. (1995) modeled the scattering and polarization 
in the central regions of NGC 1068 with a model that includes both electron
and dust scattering (Young et al. 1995). 
They could not get a fit to their NE positions unless they included electron
scatterers as well as dust in the dust cloud at the location of the northeast knot.
Packham et al. (1997) also modeled the scattered emission from the northeast knot
including both dust and electron scattering and found an acceptable fit.
Electrons are certainly present, given the ionization implied by the 
{\it HST} ionized line images 
(e.g., Capetti et al. 1997a; Bruhweiler et al. 2001; Thompson et al. 2001)
and optical spectropolarimetry (Miller et al. 1991; Inglis et al. 1995).
This ionization is from photons from the AGN and not from shocks from the
radio lobes (Bruhweiler et al. 2001).
It is probable that both electrons and dust contribute to the NIR scattered light 
from these non-nuclear regions.

\section{Summary}

We have used {\it HST} NICMOS Camera 2  
to observe the polarized light in the center of NGC 1068 at 2 \micron\ 
with a diffraction-limited PSF of $0.2''$ FWHM.
These results are one more example of the value of high-spatial-resolution observations
in disentangling complex sources, here the nucleus of one of the most important AGNs.
Since the clouds emitting the polarized light measured in this study see the AGN directly,
future work with a similar high-spatial-resolution instrument might be
to observe any nuclear activity through its echos in the polarized flux
as a function of nuclear distance (light travel time)
(e.g., Gallimore et al. 2001). 

The nucleus is an intense unresolved source, polarized at a level of 6\% 
with a position angle of $122^\circ$.
It is likely that the polarization is due to dichroic absorption in the overlying
molecular cloud.

There are two polarized lobes extending up to $8''$ northeast and southwest of the nucleus.
The polarized flux in both lobes is very clumpy, with the maximum polarization occurring
in the southwest lobe at a level of 17\% when smoothed to $0.23''$ resolution.
The southwest lobe is probably located behind the plane of the galaxy because
it cannot be detected at optical wavelengths in spite of the fact that 
it is much brighter than the northeast lobe at 2 \micron\ in polarized flux.
The perpendiculars to the polarization vectors in these two lobes all point back to the
intense unresolved source within one $0.076''$ Camera 2 pixel, 
thereby confirming that this is the illuminating source of the scattered light
and therefore the probable AGN central engine.

Features in the polarized lobes include a gap at about $1''$ between the nucleus
and the southwest lobe and a ``knot'' of emission about $4.5''$ northeast of the nucleus.
Both features had been discussed by previous authors, but they are much better defined
with the high spatial resolution of NICMOS.
The gap is sufficiently deep that we suggest that it is due to a lack of scatterers
rather than absorption by the overlying molecular cloud 
seen in CO by Schinnerer et al. (2000).  
The polarized flux images were aligned with the radio map of Wilson \& Ulvestad (1983)
by assuming that the intense point source is located at the same coordinates
as the radio source S1 (it is within $0.1''$ of S1 according to the measurements
of Capetti et al. 1997b and Thompson et al. 2001). 
The northeast knot and the north cloud  
may be the side of a molecular cloud that is facing the nucleus.
This cloud may be constraining the northeast radio lobe at the head
of the radio synchrotron-emitting jet.

Polarized fluxes were measured for positions $2.5''$ and $5.0''$ NE of the nucleus
in a $1.3 \times 2.7''$ aperture to correspond to the positions measured by
Inglis et al. (1995) from  0.43 to 0.69 \micron.
(The nucleus position was not considered because of the severe effects of extinction 
at the shorter wavelengths and excess polarized flux from dichroic absorption at 2 \micron.)
The ratio of 2 \micron\ flux to 0.6 \micron\ flux is $\sim 2.2$ at both positions.
This could indicate that the same scattering processes obtain at both $2.5''$ and $5.0''$
from the nucleus.
It also shows that the intrinsic spectrum of the NGC 1068 nucleus increases strongly
from optical wavelengths to NIR wavelengths, 
making it difficult to determine the wavelength dependence of the cross sections 
of the materials doing the scattering,
since the intrinsic spectrum of the source of the scattered light ---
the AGN central engine --- is not known.

\section{Acknowledgements}

We thank A. Wilson for the 4.9 GHz radio map used in the Fig. 3 overlay,
M. Rieke for the dark correction files, 
R. Thompson for providing a preprint, 
J. Dotson and M. Rabbette for reading the manuscript, and
E. Stobie for producing such excellent image manipulation software at Steward Observatory.
We also thank the referee for his comments and suggestions, 
which we have used to improve the presentation.
JPS and ASBS acknowledge support from 
NASA/Ames Research Center Research Interchange grants NCC2-647 and NCC2-1134
to the SETI Institute.

\clearpage

\begin{figure}
\epsscale{0.8}
\plotone{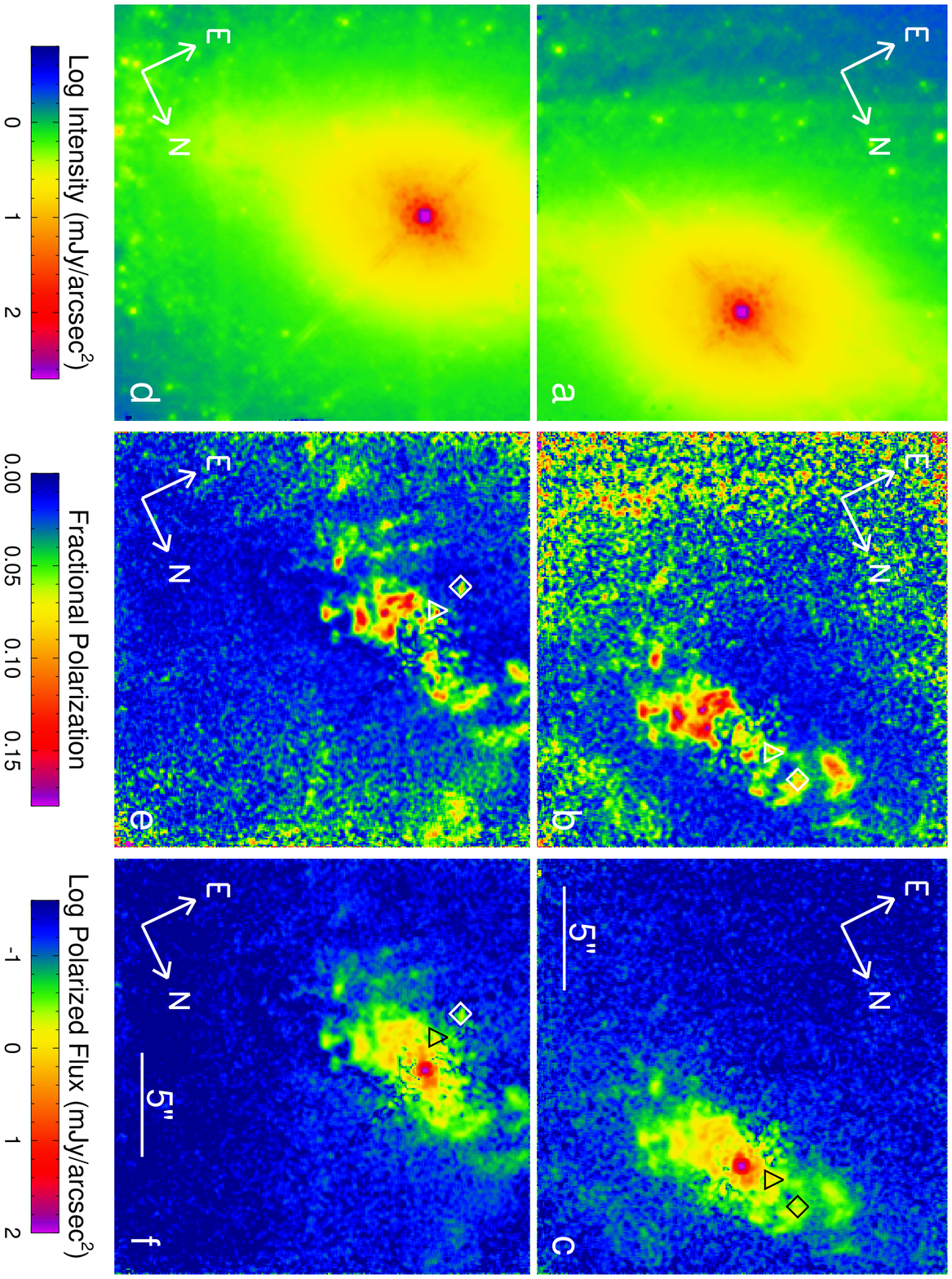}
\caption{The logarithm of the total intensity, the fractional polarization,
and the logarithm of the polarized intensity at 2 \micron\ are plotted for both visits.
The intensities are scaled so that the nucleus is saturated.
The large diamond is the POL0L ghost and the large triangle is the POL120L ghost.
Figure 1a-c: Visit 1.
Figure 1d-f: Visit 2. The Visit 2 images have been rotated by $90^\circ$ so that they have 
approximately the same orientation as the Visit 1 images.
}
\end{figure}

\begin{figure}
\plotone{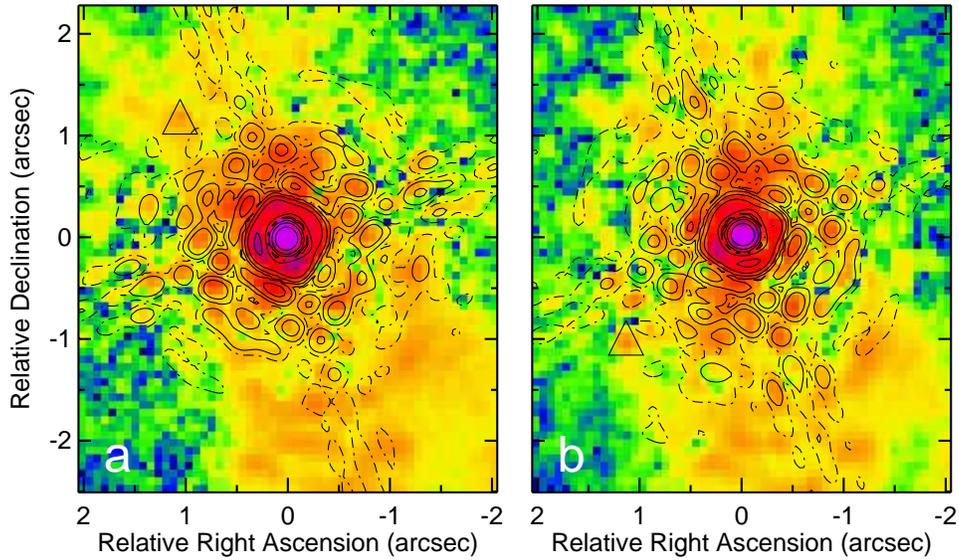}
\caption{The unsmoothed log polarized flux images (same flux scale as Fig.~1) 
from the two visits are overlaid with contours from the TinyTIM PSF calculation.
The contour levels are at 0.00032 (dot-dashed), 0.001, 0.0032, 
0.01, 0.032, 0.1, and 0.32 of the peak.
The ghost in the POL120L filter is marked by a triangle.
Both images have been rotated so that North is Up.
Features of interest that are not due to the PSF include the bright unresolved nucleus 
(magenta), 
the ridge of polarized flux (red) extending about $1''$ due north of the nucleus,
the gap of low polarized flux (green and yellow) about $1''$ south of the nucleus,
and the other regions of polarized flux (yellow and orange) that lie outside
of any PSF contour, especially the bands of polarized flux about $2''$ south of the nucleus
in the SW Lobe.
The areas that appear to be polarized flux WNW and ESE of the nucleus are the {\it HST} 
diffraction spikes from the nucleus.
a: Visit 1. b: Visit 2.
}
\end{figure}

\begin{figure}
\plotone{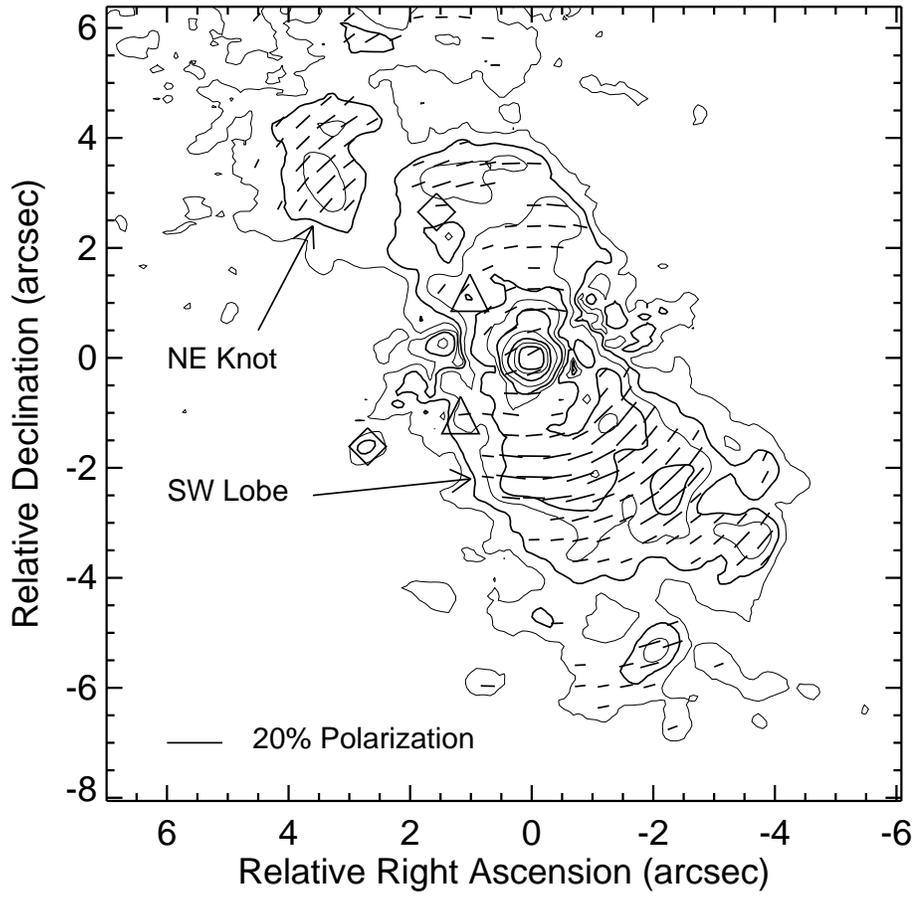}
\caption{Contours of the polarized intensity, smoothed by 5 pixels ($0.38''$),
are plotted, along with the polarization vectors. North is up. 
The contour levels are 0.08, 0.16, 0.32, 0.64, 1.28, 2.56, 5.12, 10.24, 20.48, 
and 40.96 mJy/square arcsec.
The unmarked maximum intensity is 76.3 mJy/square arcsec.
The large diamonds mark the POL0L ghost and the large triangles mark the POL120L ghost.
}
\end{figure}

\begin{figure}
\plotone{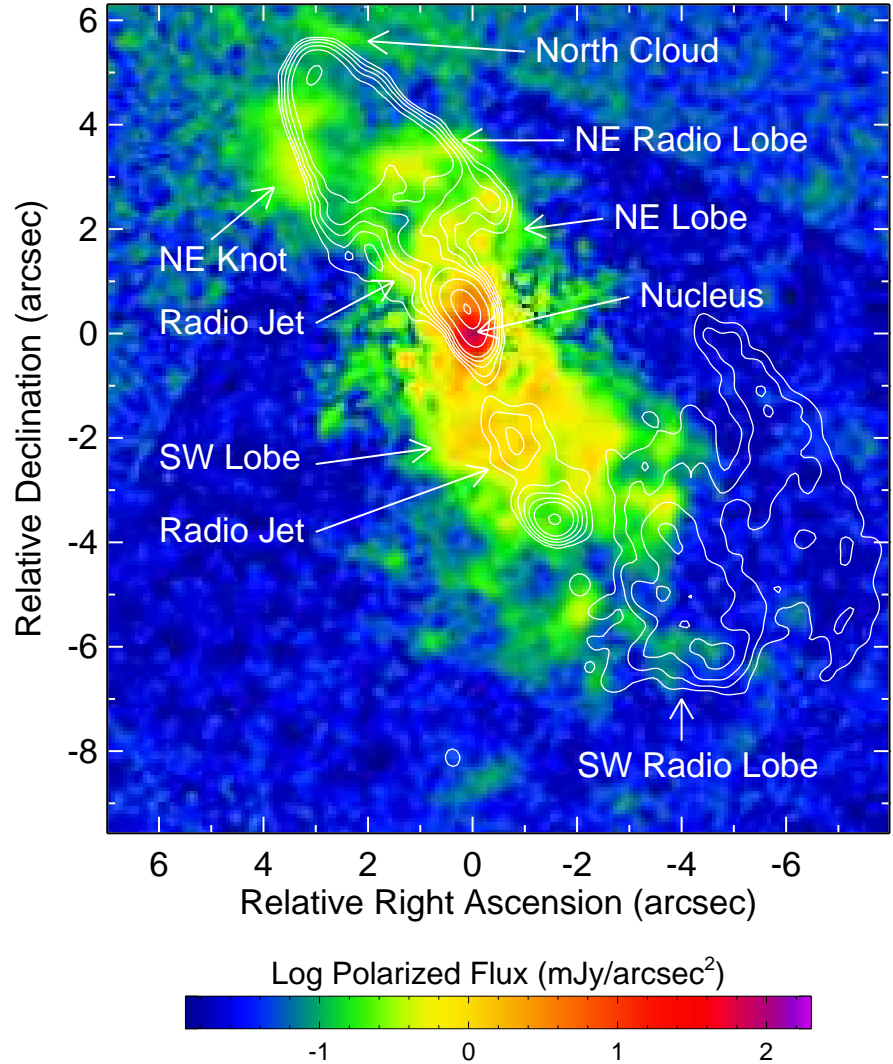}
\caption{The combined polarized flux images from the two visits are 
overlaid with contours of the 4.9 GHz radio measurements of Wilson \& Ulvestad (1983).
North is up. 
The radio contour levels are 0.0005, 0.001, 0.002, 0.004, 0.008, 0.032, 0.128, 
and 0.256 Jy beam$^{-1}$.
The images are registered by assuming that the source S1 (Muxlow et al. 1996;
Gallimore et al. 1996b)
is the AGN nucleus and is at the location of the intense NIR peak.
At the resolution of the radio image ($0.4'' \times 0.4''$),
the radio peak is dominated by radio source C, $0.29''$ north of source S1 
(Roy et al. 1998), and source S1 is not apparent.
}
\end{figure}

\clearpage

\begin{deluxetable}{lcccccc}
\tablecaption{Photometry of NGC 1068. \label{tbl-1}}
\tablewidth{0pt}
\tablehead{
\colhead{Location} & \colhead{Aperture} & \colhead{PSF\tablenotemark{a}} & \colhead{Measured} & \colhead{PSF-Corrected} & \colhead{Polarized} & \colhead{PSF-Corrected}
\\[.2ex]
\colhead{} & \colhead{arcsec} &\colhead{} &  \colhead{Flux (Jy)} & \colhead{Flux (Jy)} & \colhead{Flux(mJy)} & \colhead{Polarized Flux (mJy)}
}
\startdata
Nucleus   &   0.209\tablenotemark{b} & $0.531$ & $0.188 \pm 0.001$\tablenotemark{c} & $0.353 \pm 0.002$ & $11 \pm 2$ & $20 \pm 4$\\
Nucleus   &   0.500\tablenotemark{d} & $0.831$ & $0.315 \pm 0.001$   & $0.379 \pm 0.001$ & $18 \pm 4$ & $21 \pm 4$\\
Nucleus   &   1.000\tablenotemark{d}  & $0.908$ & $0.388 \pm 0.001$   & $0.427 \pm 0.002$ & $22 \pm 4$ & $24 \pm 4$\\
NE Knot  & $1.9 \times 1.9$\tablenotemark{e} &       &      &     & $0.74 \pm 0.13$ & \\
\enddata
\tablenotetext{a}{Fraction of PSF within the aperture calculated by TinyTIM}
\tablenotetext{b}{Radius of circular aperture corresponding to the first AIRY dark ring}
\tablenotetext{c}{Errors calculated by dividing the difference between the visits by 2}
\tablenotetext{d}{Radius of circular aperture}
\tablenotetext{e}{Size of square aperture}
\end{deluxetable}
 
\end{document}